\documentclass[11pt]{article}
\usepackage{moriond,graphicx,bm}

\def\Journal#1#2#3#4{{#1} {\bf #2}, #3 (#4)}

\def\PRD{{\em Phys. Rev.} D}

\def\RMP{\em Rev. Mod. Phys.}

\def\be{\begin{equation}}
\def\ee{\end{equation}}
\def\bea{\begin{eqnarray}}
\def\eea{\end{eqnarray}}

\begin{document}
\vspace*{4cm}
\title{CHARM AND QCD AT CLEO-III AND CLEO-c}

\author{J. NAPOLITANO}

\address{
Department of Physics, Rensselaer Polytechnic Institute, 110 Eighth Street, Troy, NY 12180 USA\\
and\\
Laboratory for Elementary Particle Physics, Cornell University, Ithaca, NY 14853 USA}

\maketitle
\abstracts{
We present recent results on charm physics from CLEO-III, emphasizing QCD and hadronic structure. These include the decay $\Xi_c^0\to pK^-K^-\pi^+$ and the form factor for $D^0\to\pi^-e^+\nu_e$. We also discuss upcoming measurements with CLEO-c, including dramatic improvements in the  $D^0\to\pi^-e^+\nu_e$ form factor, and glueball spectroscopy in $J/\psi\to\gamma X$.
}

\section{The CLEO Collaboration}

The CLEO collaboration has been in existence for over 25 years~\cite{Karl}. For the bulk of its history, CLEO has taken data on or near the $\Upsilon(4S)$ resonance, studying $B$ mesons. Charm, produced mainly in the continuum, has also been a strong focus of the CLEO program. This talk outlines two of the most recent results from this ongoing study of charm physics.

With the impressive onset of physics results from the $B$ factories Belle and BaBar, CLEO has turned its attention to specifically studying charm in the particularly clean environment encountered at lower energies. This new thrust, including upgrades to the accelerator, is called CLEO-c~\cite{Briere:2001rn}. This talk also discusses how one of our recent charm results will be improved in CLEO-c, as well as a new avenue of investigation into strong (i.e. non-perturbative) QCD, namely production of glueballs in radiative $J/\psi$ decay.

\section{The decay {\boldmath $\Xi_c^0\to pK^-K^-\pi^+$} from CLEO-III}

The $\Xi_c^0$, the lightest $csd$ baryon, naturally decays to final states such as $\Xi^-\pi^+$ where the $W^+$ from the $c\to s$ transition materializes as an external $\pi^+$. CLEO has measured~\cite{basit} the decay $\Xi_c^0\to pK^-K^-\pi^+$, including a determination of the final state fraction which is $pK^-K^{\star0}(892)$ which has no external $\pi^+$ (or $K^+$). Such decays are therefore ``color suppressed", and their rate relative to, say, $\Xi^-\pi^+$ are of specific interest.

\begin{figure}
\includegraphics[width=75mm,height=75mm]{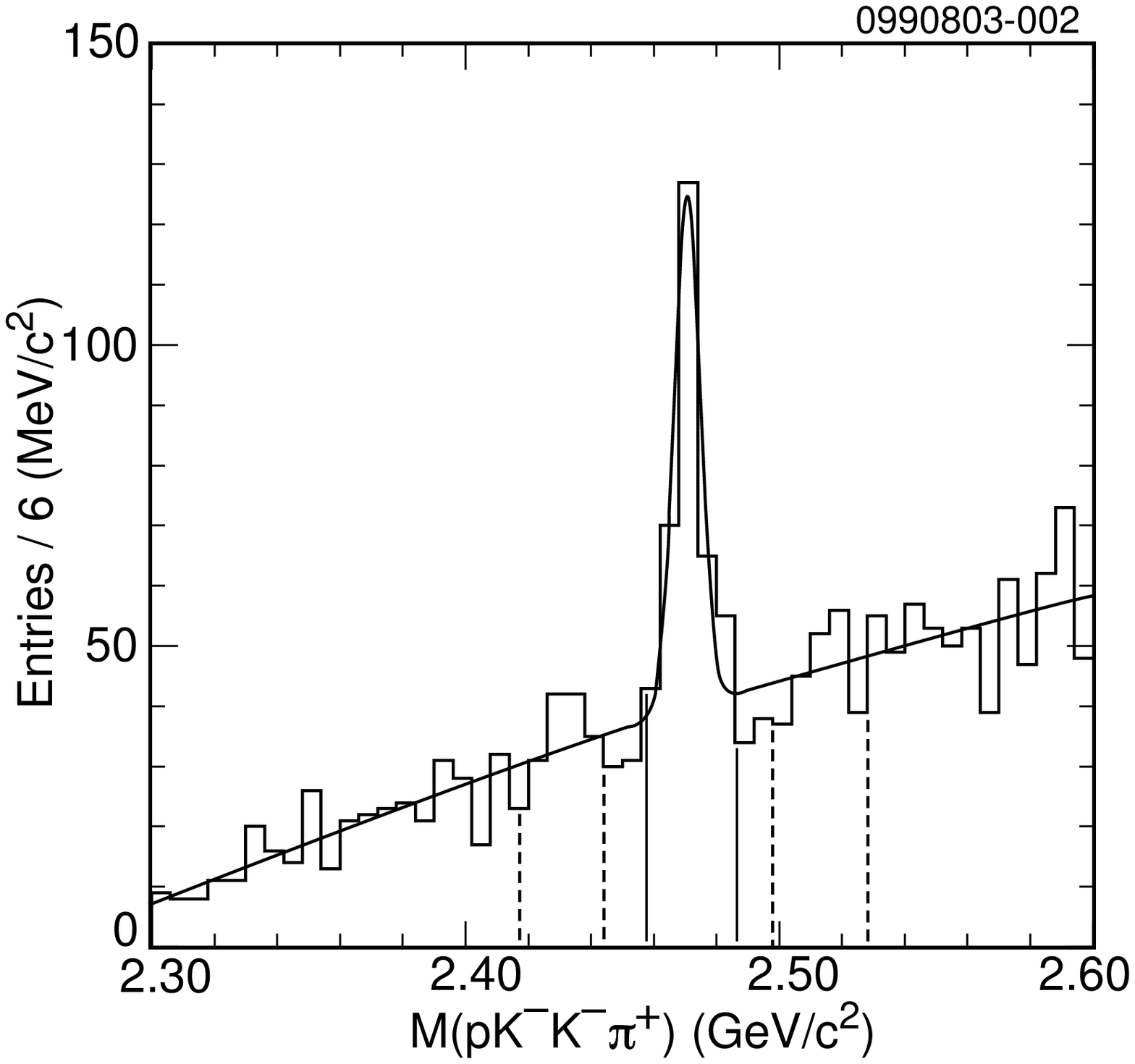}\hfill
\includegraphics[width=75mm,height=75mm]{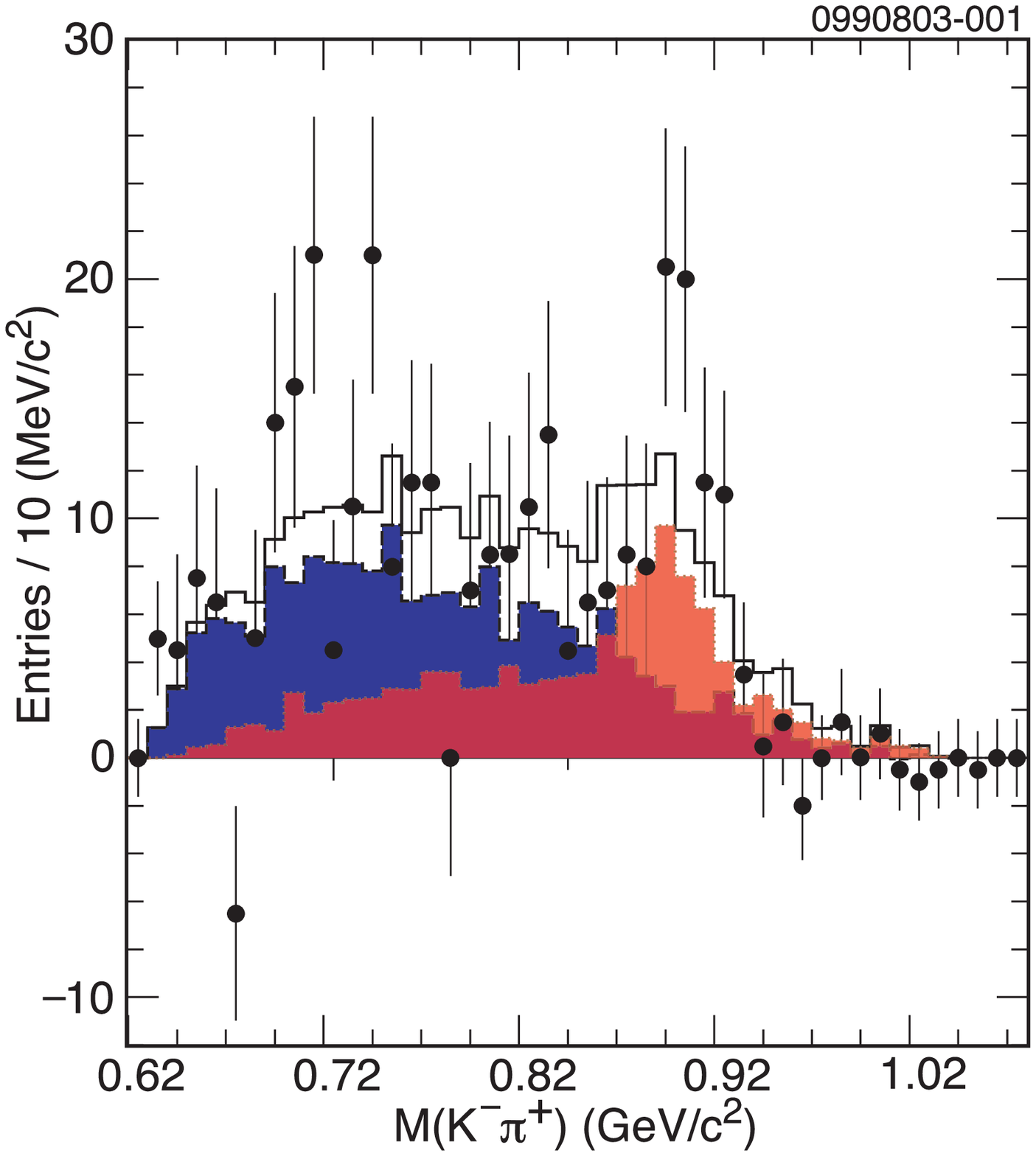}
\caption{Observation of $\Xi_c^0\to pK^-K^-\pi^+$ including the $\Xi_c^0\to pK^-K^{\star0}(892)$ component.
\label{fig:Xi_cDK}}
\end{figure}

Our signal for the $pK^-K^-\pi^+$ final state is shown in Fig.~\ref{fig:Xi_cDK},
as well as the sideband subtracted $K^-\pi^+$ mass distribution (with two entries per event). The $\Xi^-\pi^+$ is not shown, but also observed. We determine ${\cal B}(\Xi_c^0\to pK^-K^-\pi^+)/{\cal B}(\Xi_c^0\to\Xi^-\pi^+)=0.35\pm0.06\pm0.03$ and ${\cal B}(\Xi_c^0\to pK^-K^-\pi^+; {\rm No}~\bar{K}^\star)/{\cal B}(\Xi_c^0\to\Xi^-\pi^+)=0.21\pm0.04\pm0.02$.

\section{The form factor for {\boldmath $D^0\to\pi^-e^+\nu_e$}}

Using excellent $K/\pi$ discrimination and a careful treatment of systematic uncertainties in the CLEO-III detector, we have carried out the first measurement of the form factor shape (as a function of $q^2$) for $D^0\to\pi^-e^+\nu_e$. This is part of a comprehensive study of semileptonic $D^0$ decay to a single $K^-$ or $\pi^-$. This is a challenging measurement using $D^0$ from $D^{*+}\to\pi^+_{\rm slow}D^0$ produced in $e^+e^-$ annihilation at $\Upsilon(4S)$ energies. It is also a measurement that we are preparing to take up in CLEO-c, where kinematic separation of $K^-$ and $\pi^-$ final states becomes possible.

\subsection{CLEO-III}

\begin{figure}
\begin{minipage}{75mm}
\includegraphics[width=75mm]{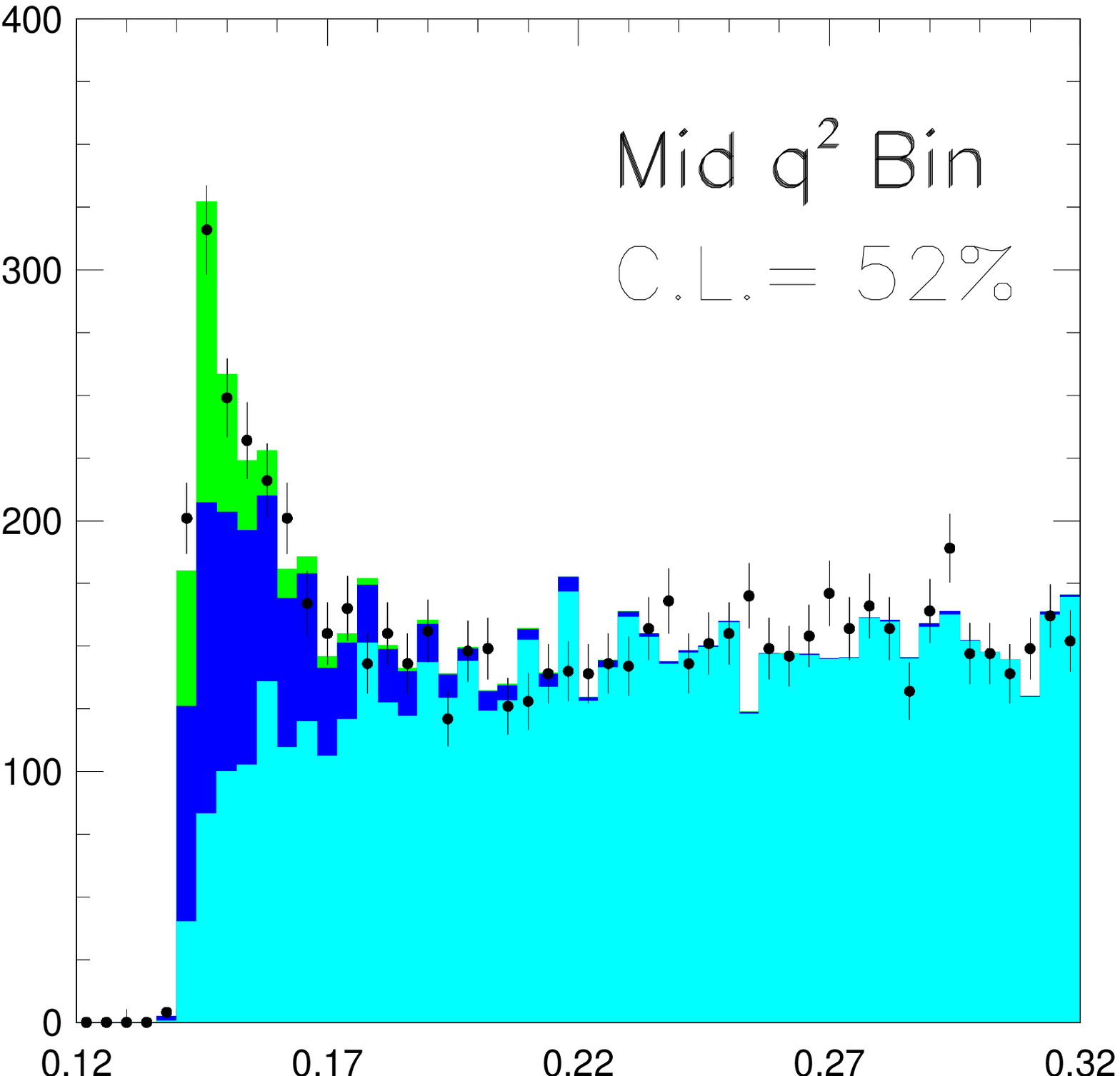}
\centerline{$\Delta M$ (GeV$/c^2$)}
\end{minipage}\hfill
\begin{minipage}{75mm}
\includegraphics[width=75mm]{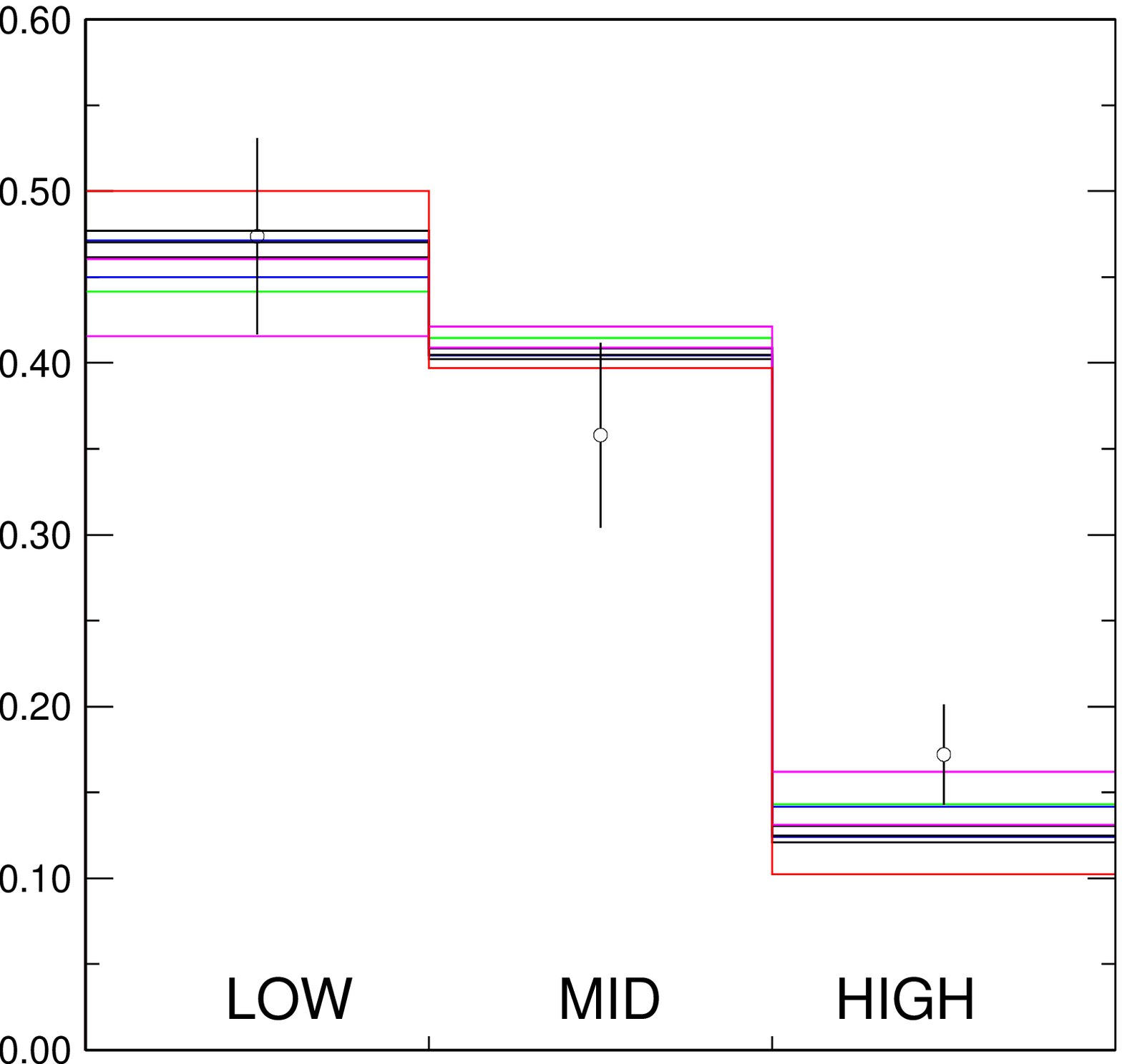}
\centerline{$q^2$ (GeV$^2/c^2$)}
\end{minipage}
\caption{Measurement of the form factor shape for $D^0\to\pi^-e^+\nu_e$ in CLEO-III. The left plot histograms $\Delta M=M\left(\pi^+_{\rm slow}\pi^-e^+\nu_e\right)-M(D^0)$ and peaks in the region of the signal. Data points with error bars are superimposed on top of the fit, whose components are plotted as shaded histograms. The components are the non-peaking random backgrounds, the peaking background from $K^-$ misidentified as $\pi^-$, and the signal, which is the smallest of the three and plotted on top of the others. On the right we show the observed rate (uncorrected for detector efficiency)
in three $q^2$ bins, and compared to various models.
\label{fig:D0FFIII}}
\end{figure}

Figure~\ref{fig:D0FFIII}
shows our signal for $D^0\to\pi^-e^+\nu_e$, for the middle of three $q^2$ bins, and also our result for the decay rate (normalized to unity and not corrected for detector efficiency) as a function of $q^2$. The histograms on the right show the range of various calculations of the form factor. Another result of this analysis is a new measurement of the relative branching ratios for $D^0\to\pi^-e^+\nu_e$ and $D^0\to K^-e^+\nu_e$. We find ${\cal B}(D^0\to\pi e\nu)/{\cal B}(D^0\to Ke\nu)=0.097\pm0.010\pm0.010$. Results presented here are preliminary, but final results are nearing submission at this writing.

\subsection{CLEO-c}

\begin{figure}
\begin{minipage}{90mm}
\includegraphics[width=90mm]{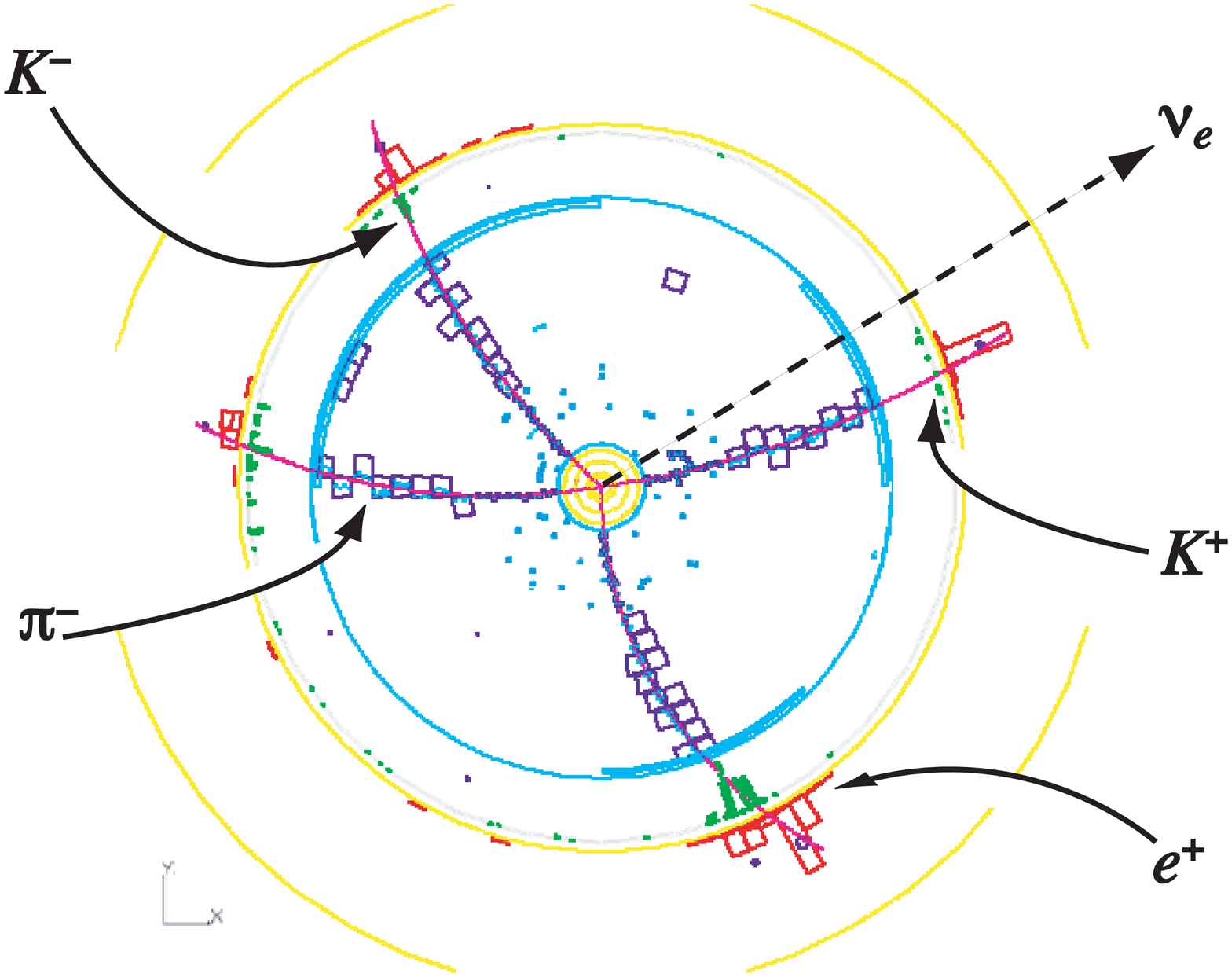}
\end{minipage}\hfill
\begin{minipage}{60mm}
\includegraphics[width=60mm]{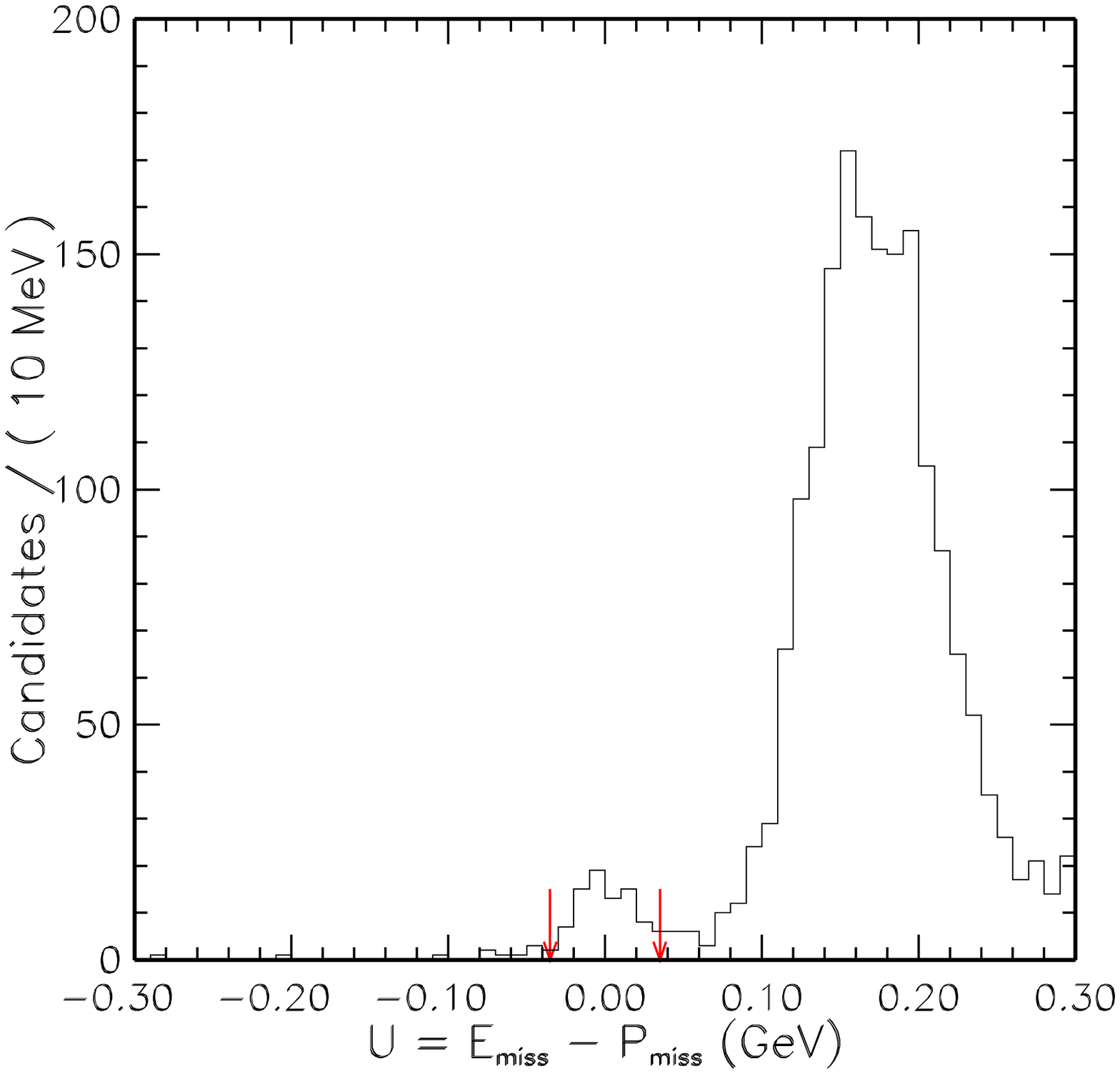}
\end{minipage}
\caption{Measuring semileptonic $D^0$ decay in CLEO-c. On the left is an event display showing $e^+e^-\to\bar{D}^0D^0$ with a $\bar{D}^0\to K^+\pi^-$ decay which serves to ``tag'' the event, and a semileptonic $D^0\to K^-e^+\nu_e$ decay. The histogram shows how the quantity $E_{\rm miss}-p_{\rm miss}$ cleanly separates the $\pi^-$ and $K^-$ semileptonic decays.
\label{fig:D0FFc}}
\end{figure}

CLEO-c will obtain a very large sample of $e^+e^-\to D^0\bar{D}^0$ events. One example from our preliminary running is shown on the left in Fig.~\ref{fig:D0FFc}.
Such events are very clean and highly constrained kinematically. On the right we show the use of those kinematic constraints to separate $D^0\to\pi^-e^+\nu_e$ from the much more dominant $D^0\to K^-e^+\nu_e$, without the need for $K/\pi$ separation through particle identification. The figure represents roughly 2\% of the eventual CLEO-c data sample.

\section{Glueball spectroscopy using {\boldmath $J/\psi\to\gamma X$} in CLEO-c}

Radiative decays of the $J/\psi$ have long been used to search for glueballs~\cite{Godfrey:1998pd}, and this data set has recently seen a great increase in statistics in some channels~\cite{Bai:2003ww}. Nevertheless, many puzzles remain. One significant problem is that glueballs are expected to mix with $q\bar{q}$ states in the $1400-1800$~MeV/$c^2$ region, and it is difficult to discern the bare components within the physical states.

CLEO-c aims to acquire $10^9$ $J/\psi$ decays, a factor of $\sim20$ over existing samples. In addition, all the resources of the CLEO detector, including excellent photon detection and particle identification, can be brought to bear on the problem. Radiative decays and glueball spectroscopy will be a keystone of the program.

These enhancements of the data sample will make some new analyses possible. One example, targeted at disentangling the glueball and $q\bar{q}$ components in the scalar mesons, has been suggested~\cite{close} and is summarized in Table~\ref{tab:DblRadDK}.
\begin{table}[t]
\begin{center}
\caption{Discrimination between different glueball/$q\bar{q}$ mixing scenarios, using radiative decay of scalar mesons.\label{tab:DblRadDK}}
\vspace{0.4cm}
\begin{tabular}{|l|rrr|rrr|r|}
\hline
\multicolumn{7}{|c|}{Radiative Decay Widths in keV} & {$\Gamma_{\rm Tot}$} \\
\hline
& \multicolumn{3}{c|}{$f_0\to\gamma\rho(770)$} & \multicolumn{3}{c|}{$f_0\to\gamma\phi(1020)$}
& \multicolumn{1}{c|} {MeV} \\
\hline
State & L & M & H & L & M & H & \\
\hline
$f_0(1370)$ &   443 & 1121 & 1540 & 8 & 9 & 32        & {$\sim$300} \\
$f_0(1500)$ & 2519 & 1458 &   476 & 9 & 60 & 454    & {109} \\
$f_0(1710)$ &      42 &     94 &   705 & 800 & 718 & 78 & {125}\\
\hline
\end{tabular}
\end{center}
\end{table}
The labels ``L'', ``M'', and ``H'' refer to the case where the bare glueball is lighter than the isoscalar $u\bar{u}+d\bar{d}$ component, in between this and, or heavier than, the $s\bar{s}$ component. The relative radiative decays of the physical scalar mesons $f_0(1370)$, $f_0(1500)$, and $f_0(1710)$ are clearly sensitive measures of these cases. The total widths of these states are also listed, showing that the branching ratios are between $\sim10^{-4}$ and $\sim10^{-2}$. Given a branching ratio $J/\psi\to\gamma f_0$ of $\sim10^{-3}$, one expects between 100 and $10^4$ events for these double radiative decays. This should be enough to experimentally examine Table~\ref{tab:DblRadDK}.

\section*{Acknowledgments}
I gratefully acknowledge the kind hospitality of the Laboratory for Elementary Particle Physics, and Cornell University during this past year. This work was supported by a grant from the National Science Foundation.


\section*{References}
\begin{thebibliography}{99}

\bibitem{Karl}K.\ Berkelman, ``A Personal History of CESR and CLEO'', World Scientific Publishing (2004)

\bibitem{Briere:2001rn} R.~A.~Briere {\it et al.},
``{CLEO}--c and {CESR}--c: A New Frontier of Weak and Strong Interactions,''
CLNS-01-1742, {\sf http://www.lns.cornell.edu/public/CLEO/spoke/CLEOc/}

\bibitem{basit}I.\ Danko {\it et al.}, \Journal{\PRD}{69}{052004}{2004}

\bibitem{Godfrey:1998pd} S.~Godfrey and J.~Napolitano, \Journal{\RMP}{71}{1411}{1999}

\bibitem{Bai:2003ww} J.~Z.~Bai {\it et al.}  [BES Collaboration], \Journal{\PRD}{68}{052003}{2003}

\bibitem{close}F.\ E.\ Close {\it et al.}, \Journal{\PRD}{67}{074031}{2003}

\end{thebibliography}
\end{document}